\documentclass[a4paper]{jpconf}
\usepackage{graphicx}

\usepackage{amsmath}
\usepackage{amssymb,epsfig}

\usepackage{stmaryrd,wasysym,upgreek,latexsym}
\usepackage{lscape}
\usepackage[vflt]{floatflt}
\usepackage{afterpage}

\newcommand{\beq}{\begin{equation}}
\newcommand{\eeq}{\end{equation}}
\newcommand{\beqa}{\begin{eqnarray}}
\newcommand{\eeqa}{\end{eqnarray}}
\newcommand{\noi}{\noindent}
\newcommand{\om}{\Omega}
\newcommand{\tom}{\tilde \Omega}

\newcommand{\resetequ}{\setcounter{equation}{0}}
\renewcommand{\thesection}{\Roman{section}}

\begin{document}
\title{Overview of the parametric representation of renormalizable non-commutative field theory}

\author{Adrian Tanasa}

\address{Laboratoire de Physique Th\'eorique, \\
b\^at. 210, CNRS UMR 8627,\\
Universit\'e Paris XI, 91405, Orsay Cedex, France\\
Dep. Fizica Teoretica,\\
Institutul de Fizica si Inginerie Nucleara H. Hulubei\\
 P. O. Box MG-6, 077125 Bucuresti-Magurele, Romania}

\ead{adrian.tanasa@ens-lyon.org}

\begin{abstract}
We review here the parametric representation of Feynman amplitudes of
 renormalizable non-commutative quantum field models.
\end{abstract}

\section{Introduction}
\resetequ

In the quest for quantification of gravity, non-commutative
geometry (see \cite{book-connes}) is nowadays one of the most appealing concepts. Non-commutative quantum
field theory (NCQFT) - for some general reviews see \cite{DN} or \cite{Szabo}
- is also
an interesting candidate for New Physics beyond the Standard Model. Moreover,
let us also note that NCQFT can be seen as an effective limit of string theory
\cite{CDS, SW}).

In a completely different framework, non-commutativity can be important for
the understanding of physics  in the presence of a
background field (like is the case for example for the fractional quantum Hall
effect \cite{Suss, Poly, HellRaams}).

Because of a new type of divergence - the UV/IR mixing - NCQFTs were thought to
be non-renormalizable. The situation changed for scalar fields with the
Grosse-Wulkenhaar model, model which modifies the kinetic part of the
action. This modification takes into consideration a symmetry already present
at the level of the interaction term, the Langman-Szabo duality \cite{ls},
duality between the IR and UV regions). 
This model was proven, by different methods, to be renormalizable to any order in
perturbation theory \cite{GW1, GW2, GW3, RVW, GMRV}. The parametric
representation for it was then implemented \cite{param1}. This was the starting
point for the proof of  dimensional renormalization \cite{dimreg}. Recently,
this modification of the kinetic part of the action was proven \cite{ultimul} to be also
related to the spectral action principle (for last advances to this issue see
\cite{spectral}).

The Grosse-Wulkenhaar model was moreover proven to have a better theoretical behavior
with respect to the commutative $\phi^4$ model. Indeed, in \cite{beta1,
  beta23} and \cite{beta} was proven that this model does not present a Landau
ghost; let us recall that this was not the case for the commutative model. 

Another improvement with respect to commutative scalar quantum field theory is
that a constructive version (for a general review see \cite{carte}) is within
reach \cite{constructiva1, constructiva2}.

An analyze of the vacuum states of the model as well as an investigation of
spontaneous symmetry breaking and the Goldstone theorem for a corresponding
linear sigma model were also realized \cite{goldstone}.

Moreover, a second type of renormalizable NCQFT models - the {\it covariant}
models - exists \cite{V}. This
classification is done with respect to the form of the propagator in position
space. This second category of NCQFTs contains the non-commutative Gross-Neveu
model and the Langmann-Szabo-Zarembo (LSZ) model \cite{lsz}. The parametric
representation for the latter was also obtained \cite{param2}. One can report
himself to \cite{sefu} or \cite {raimar} for some general review of renormalizable NCQFTs.

For both types of models the Mellin representation of the Feynman amplitudes
was implemented \cite{mellin}. Furthermore, a Hopf algebra structure associated
to this new type of renormalization was defined \cite{hopf}.

Finally, let us also state that recent progress has been obtained in the
search of renormalizable non-commutative gauge theories \cite{ultimul, gauge1,
  gauge2, gauge3, gauge4, gauge5}.

\medskip

We present here an overview of the parametric representation for both
known types of renormalizable NCQFTs, overview based on \cite{param1} and
\cite{param2}.  The paper is structured as follows. In
the next section, a brief recall of the parametric representation for
commutative QFT is given. In the third section we introduce the two types of
models we consider here: the Grosse-Wulkenhaar model and the covariant models. The next section presents our main results, namely
the implementation of the parametric representation for both these models. Finally,
in the last section we give a few examples for some simple Feynman graphs.

\section{Parametric representation for commutative QFT}
\resetequ

The parametric representation relies on the introduction of an integral representation
on some Schwinger parameters $\alpha_\ell$ ($\ell=1,\ldots, L$, $L$ being the total number of internal lines of the graph).
One has
$$\frac{1}{p^2_\ell}=\int_0^\infty d\alpha_\ell e^{-\alpha_\ell k^2_\ell}. $$
The amplitude of a Feynman graph for a non-massif  model writes 
(see for example \cite{carte} or \cite{itz} for further details)
\beqa 
\label{as} 
{\cal A} (p) = \delta(\sum p)\int_0^{\infty} 
\frac{e^{- V(p,\alpha)/U (\alpha) }}{U (\alpha)^{2}} 
\prod_{\ell=1}^L  
d\alpha_\ell.
\eeqa
\noi
The $U$ and $V$, so called ``topological" or
``Symanzik" 
polynomials, are proven to be polynomials in the Schwinger parameters $\alpha$
and they have 
for the $\phi^4$ model 
the expressions
\beqa
\label{s1}
U &=& \sum_{\cal T} \prod_{l \not \in {\cal T}} \alpha_\ell \ , 
\eeqa
\beqa
\label{s2}
V &= &\sum_{{\cal T}_2} \prod_{l \not \in {\cal T}_2} \alpha_\ell  (\sum_{i \in
  E({\cal T}_2)} p_i)^2 \ , 
\eeqa
\noi
where $\cal T$ is a (spanning) tree of the graph and ${\cal T}_2$ is a
$2-$tree, {\it i. e.} a tree minus one of its lines. Any such $2-$tree splits
  the graph into  exactly two connected components which we denote $E({\cal
  T}_2)$ and $F({\cal T}_2)$. Moreover the sum 
$\sum_{i \in  E({\cal T}_2)} p_i$ (where $p_i$ is an external momentum) is, by momentum conservation, also
equal to $ (\sum_{i \in F(T_2)} p_i)^2$.

\section{Renormalizable NCQFT models}
\resetequ

We place ourselves in a $4-$dimensional Moyal space
\beqa
\label{2D}
[x^\mu, x^\nu]=i \Theta^{\mu \nu},
\eeqa
\noi
where the the matrix $\Theta$ is
\begin{eqnarray}
\label{theta}
  \Theta= 
  \begin{pmatrix}
0 &\theta & 0 & 0\\ 
-\theta & 0 & 0 & 0\\
0 & 0 & 0 & \theta \\
0 & 0 & -\theta & 0
  \end{pmatrix}.
\end{eqnarray}
\noi
The associative Moyal product of two functions 
$f$ and $g$ 
on the Moyal space writes
\beqa
\label{moyal-product} 
 (f\star g)(x)=\int \frac{d^{4}k}{(2\pi)^{4}}d^{4}y\, f(x+{\textstyle\frac 12}\Theta\cdot
  k)g(x+y)e^{\imath k\cdot y}\nonumber\\
\eeqa
We also consider Euclidean metric. Let us now introduce in the rest of this
section the two types
of renormalizable non-commutative models.

\subsection{The Grosse-Wulkenhaar model}


Note that the results  established in the sequel hold for orientable models
(that is interactions $\bar \phi \star \phi \star \bar \phi \star \phi$). This corresponds to a Grosse-Wulkenhaar
model of a complex scalar field
\beqa
\label{lag}
S_{GW}=\int d^4 x \left(\frac{1}{2} \partial_\mu \bar \phi
\star \partial^\mu \phi +\frac{\Omega^2}{2} (\tilde{x}_\mu \bar \phi )\star
(\tilde{x}^\mu \phi ) 
+  \bar \phi \star \phi \star \bar \phi \star \phi \right)
\eeqa
where 
\beqa
\label{tildex}
\tilde{x}_\mu = 2 (\Theta^{-1})_{\mu \nu} x^\nu
\eeqa
\noi
This action leads to the following  propagator from a point $x$ to a point $y$:
\begin{equation}
\label{propa1}
C(x,y)=\int_0^\infty \frac{\tilde \Omega d\alpha}{[2\pi\sinh(\alpha)]^{2}}
e^{-\frac{\tilde \Omega}{4}\coth(\frac{\alpha}{2})(x-y)^2-
\frac{\tilde \Omega}{4}\tanh(\frac{\alpha}{2})(x+y)^2}\; .
\end{equation} 
Let us now introduce the {\it short} and {\it long variables}:
\beqa
\label{def-uv}
u=\frac{1}{\sqrt 2} (x-y),\ v=\frac{1}{\sqrt 2} (x+y).
\eeqa
and
\beqa
\label{t}
t_\ell = {\rm tanh} \frac{\alpha}{2}.
\eeqa
The propagator \eqref{propa1} becomes
\begin{equation}
\label{propa}
C(x,y)=\int_0^\infty \frac{\tilde \Omega d\alpha}{[2\pi\sinh(\alpha)]^{2}}
e^{-\frac{\tilde \Omega}{2}\frac{1}{t_\ell} u^2-
\frac{\tilde \Omega}{2}t_\ell v^2}\; .
\end{equation}

\subsection{The covariant models}

Amongst this type of models one has, as already stated in the introduction, the
non-commutative Gross-Neveu model and the LSZ model. The
results we establish in the sequel hold for the latter but they can be
however extended for the Gross-Neveu model also. The
LSZ action writes:
\beqa
\label{lag2}
S_{LSZ}=\int d^4 x \left(\frac 12 (\partial_\mu \bar \phi - i \Omega \tilde x_\mu\phi)
\star (\partial^\mu  \phi -i \Omega  \tilde x^\mu \phi)
 + \bar \phi\star\phi\star\bar \phi\star\phi \right).
\eeqa
This action leads to the propagator
\beqa
\label{propa2}
C(x,y)=2 \int_0^1 d t_\ell \frac{\tilde \Omega(1-t_\ell^2)}{(4\pi
  t_\ell)^2}e^{-\frac 12 \tom \frac {1+ t_\ell^2}{2 t_\ell}u^2 + i \tom  u\wedge v},
\eeqa
where
$$ u\wedge v = u_1 v_2 - u_2 v_1 + u_3 v_4 - u_4 v_3.$$

\subsection{Non-local interaction and hypermomenta}

Using the form \eqref{moyal-product} of the Moyal product, the interaction
term of both \eqref{lag} and \eqref{lag2} 
lead to the following contribution in position space
\beqa
\label{v1}
\delta (x_1^V - x_2^V + x_3^V - x_4^V)e^{2i\sum_{1\le
    i <j\le 4}(-1)^{i+j+1}x_i^V\Theta^{-1}x_j^V}
\eeqa
\noi
where $x_1^V,\ldots, x_4^V$ are the $4-$vectors of the positions of the $4$
fields incident to the respective vertex.

To any such vertex one associates 
a hypermomentum $p_V$ {\it via}
the relation
\beqa
\label{pbar1}
\delta(x_1^V -x_2^V+x_3^V-x_4^V ) 
= \int  \frac{d p_V}{(2 \pi)^4}
e^{p_V \sigma (x_1^V-x_2^V+x_3^V-x_4^V)}
\eeqa

\section{Parametric representation}
\resetequ

In the case of 
commutative QFT, one has translation invariance in position
space. As a consequence of this invariance, the first polynomial $U$ (see
section $II$) vanishes when
integrating over all internal positions. Therefore, one has to integrate over
all internal positions (which correspond to vertices) save one, which is thus
singularized. However, the polynomial is a still a canonical object, {\it
  i. e.} it does not depend of the choice of this
particular vertex.

As noticed in \cite{param1}, in the non-commutative case the translation
invariance is lost. Therefore, one can integrate
over all internal positions and hypermomenta, without vanishing of the first
polynomial. However, in order to be able to recover the commutative limit, we
also singularize a particular vertex $\bar V$; we do not integrate on its associate hypermomenta
$p_{\bar V}$. We call this particular vertex the {\it
  root}. Because there is no translation invariance, the polynomial
does  depend on the choice of the root; however the leading ultraviolet
terms do not.

\medskip

We now define the
$(L\times 4)$-dimensional incidence matrix $\e^V$ for each of the vertices
$V$. 
Since the graph is orientable (in the sense defined in the previous section) we can choose
\beqa
\label{r1}
\epsilon_{\ell i}^V= (-1)^{i+1}, \mbox { if the line $\ell$ hooks to the vertex $V$
  at corner $i$.}
\eeqa
\noi
We also put
\beqa
\label{r2}
\eta^V_{\ell,i}=\vert \epsilon^V_{\ell,i}\vert, \mbox { } V=1,\ldots, n,\, 
\ell=1,\ldots, L \mbox{ and } i=1,\ldots, 4. 
\eeqa
From \eqref{r1} and \eqref{r2} one has
\beqa
\label{r3}
\eta^V_{\ell i} = (-1)^{i+1}\epsilon_{\ell i}.
\eeqa
We now generalize the short and long variables introduced in the previous
section at the level of the whole Feynman graph:
\beqa
v_\ell&=&\frac{1}{\sqrt{2}} \sum_V \sum_i \eta^V_{\ell i} x^V_i,\nonumber\\
u_\ell&=&\frac{1}{\sqrt{2}} \sum_V \sum_i \epsilon^V_{\ell i} x^V_i.
\eeqa
\noi
Conversely, one has
$$ x^V_i= \frac{1}{\sqrt{2}}\left(\eta^V_{\ell i}v_\ell+\e^V_{\ell i}u_\ell \right).$$

\medskip

From the propagator \eqref{propa} and the vertices
contributions \eqref{v1} one is able to write the amplitude ${\cal A}_{G,{\bar
    V}}$ of the graph $G$ (with the singularized root $\bar V$) as function of
the non-commutative polynomials $HU_{G, \bar{V}}$ and $HV_{G, \bar{V}}$ as (see
\cite{param1} for details)
\beqa
\label{HUGV}
{\cal A}_{G,{\bar V}}  (x_e,\;  p_{\bar V}) = K \int_{0}^{1} \prod_{\ell=1}^L  [ d t_\ell
(1-t_\ell^2) ]
HU_{G, \bar{V}} ( t )^{-2}   
e^{-  \frac {HV_{G, \bar{V}} ( t_\ell , x_e , p_{\bar v})}{HU_{G, \bar{v}} ( t )}},
\eeqa
where  $K$ is some constant, unessential
for this calculus and by $x_e$ we mean the external positions of the graph, 
In \cite{param1} it was furthermore proved that  $HU$ and $HV$
are polynomials in the set of variables $t$.

Let us state that, even the formulas
above hold also for non-orientable graphs (that is graphs corresponding to
interactions $\bar \phi \star \bar \phi \star \phi \star \phi$), for simplicity reasons 
we restrict ourselves
to the study of polynomials for orientable graphs (that is graphs corresponding to
interactions $\bar \phi \star \phi \star \bar \phi \star \phi$, as already
mentioned in the previous section). One has 
(see again \cite{param1})
\beqa
\label{huqv12}
HU_{G, \bar{V}}=({\rm det} Q)^{\frac 14} \prod_{\ell=1}^L t_\ell
\eeqa
where
\beqa
\label{Q}
Q= A\otimes 1_D - B \otimes \sigma,
\eeqa
with $A$ a diagonal matrix and $B$ an antisymmetric matrix.

\subsection{The parametric representation for the Grosse-Wulkenhaar model}

The matrix $A$ writes 
\beqa 
\label{defmatrixa}
 A=\begin{pmatrix} S & 0 & 0\\ 0  & T & 0 \\ 0&0&0\\
\end{pmatrix},
\eeqa 
where $S$ and resp. $T$ are the two diagonal $L$ by $L$ matrices
with diagonal elements $c_\ell = \coth(\frac{\alpha_\ell}{2}) = 1/t_\ell$, 
and resp. $t_\ell$.
The last $(n-1)$ lines and columns are have $0$ entries.

The antisymmetric part $B$ is
\beqa
\label{b}
B= \begin{pmatrix}{s} E & C \\
-C^t & 0 \\
\end{pmatrix}\ 
\eeqa
\noi
with 
$$s=\frac{2}{\theta\tom}=\frac{1}{\om}$$
and 
\beqa
\label{c}
C_{\ell V}=\begin{pmatrix}
\sum_{i=1}^4(-1)^{i+1}\epsilon^V_{\ell i} \\
\sum_{i=1}^4(-1)^{i+1}\eta^V_{\ell i} \\
\end{pmatrix}\ ,
\eeqa
\noi
\beqa
\label{e}
E=\begin{pmatrix}E^{uu} & E^{uv} \\ E^{vu} & E^{vv} \\
\end{pmatrix},
\eeqa
where
\beqa
\label{ee}
E^{vv}_{\ell,\ell'}&=&\sum_V
\sum_{i,j=1}^4  (-1)^{i+j+1} \omega(i,j)\eta_{\ell i}^V\eta_{\ell' j}^V,
\nonumber\\
E^{uu}_{\ell,\ell'}&=&\sum_V
\sum_{i,j=1}^4  (-1)^{i+j+1} \omega(i,j)\epsilon_{\ell i}^V\epsilon_{\ell' j}^V,
\nonumber\\
E^{uv}_{\ell,\ell'}&=&\sum_V
\sum_{i,j=1}^4  (-1)^{i+j+1} \omega(i,j)\epsilon_{\ell i}^V\eta_{\ell'
  j}^V. 
\eeqa 
\noi
Note that $\omega$ is the antisymmetric matrix for whom $\omega(i,j)=1$ if $i<j$.
Finally, in order to have the integer expression \eqref{c} of the matrix $C$
we have rescaled by $s$ the hypermomenta $p_V$. 
We also define the integer entries matrix:
\beqa
\label{b'}
B'= \begin{pmatrix} E & C \\
-C^t & 0 \\
\end{pmatrix}\ .
\eeqa

In \cite{param1} it was also proven that
\beqa
\label{qm}
{\rm det} Q= ({\rm det} M)^4
\eeqa
where
\beqa
\label{M}
M= A+B
\eeqa
and thus \eqref{huqv12} becomes:
\beqa
\label{hugvq2}
HU_{G, \bar{V}}={\rm det} M \prod_{\ell=1}^L t_\ell
\eeqa
\noi
 
%

Let now $I$
and resp. $J$ be two subsets of $\{1,\ldots,L\}$, of cardinal $\vert I \vert$
and resp. $\vert J \vert$. Moreover, let
\beqa
\label{kij}
k_{I,J} = \vert I\vert+\vert J\vert - L - F +1
\eeqa
and $n_{I J}=\mathrm{Pf}(B'_{\hat{I}\hat{J}})$, the Pfaffian of the  matrix
$B'$  
with deleted lines and columns $I$ among the first $L$ indices 
(corresponding to short variables $u$) and $J$ among the next $L$ 
indices (corresponding to long variables $v$).
 
The specific form \eqref{defmatrixa}
allows one to write the polynomial $HU$ as a sum of positive terms:
\beqa
\label{suma}
HU_{G,{\bar V}} (t) &=&  \sum_{I,J}  s^{2g-k_{I,J}} \ n_{I,J}^2
\prod_{\ell \not\in I} t_\ell \prod_{\ell' \in J} t_{\ell'}\ .
\eeqa
where $g$ is the genus of the graph.

In \cite{param1}, non-zero {\it leading terms} ({\it i. e.} terms which have
the smallest global degree in the $t$ variables) were identified. 
These terms are dominant in the UV regime. 
Some of them 
correspond to subsets $I=\{1,\ldots, L\}$ and $J$  { admissible},  where by
{\it admissible} we mean that 
\begin{itemize}
\item it contains a tree $\tilde \cal T$ in the dual graph,
\item its complement contents a tree $\cal T$ in the direct graph.
\end{itemize}

In this case,
$$ n_{I,J}=2^{2g-k_{J}}$$
where $k_J$ is nothing but $k_{I,J}$ given by \eqref{kij} with $|I|=L$.
This allows us to set a lower limit on the polynomial $HU$:
\beqa
\label{limit}
HU_{G,\bar V}\ge 
\sum_{J\, admissible} (2s)^{2g-k_{J}} \prod_{\ell\in J} t_\ell,
\eeqa
This is the main result obtained in \cite{param1}.

\subsection{Parametric representation for the Langman-Szabo-Zarembo Model}

As stated above, the different form of the propagator \eqref{propa2} (with
respect to the Grosse-Wulkenhaar propagator \eqref{propa}) leads to a series
of changes in the parametric representation for the covariant models, changes which
we now list (see \cite{param2} for further details).
The diagonal matrix $A$ writes 
\beqa 
\label{defmatrixa2}
 A=\begin{pmatrix} S & 0 & 0\\ 0  & 0 & 0 \\ 0&0&0\\
\end{pmatrix}
\eeqa 
where $S$ is a $L-$dimensional diagonal matrix with elements $\frac{1+t_\ell^2}{2t_\ell}$.

The antisymmetric part $B$ has the same form as \eqref{b}, the only difference
being in:
\beqa
\label{ee2}
E^{uv}_{\ell,\ell'}=\sum_V
\sum_{i,j=1}^4  (-1)^{i+j+1} \omega(i,j)\epsilon_{\ell i}^V\eta_{\ell'
  j}^V+ 2 \Omega \delta_{\ell \ell'}. 
\eeqa 
\noi

Let now $K$
be a subset of $\{1,\ldots,L\}$. Let
\beqa
\label{kij2}
k_{K} = \vert K\vert- L - F +1
\eeqa
and $n_{K}=\mathrm{Pf}(B'_{\hat{K}})$, the Pfaffian of the  matrix
$B'$  
with deleted lines and columns $K$ among the first $L$ indices 
(corresponding to short variables $u$).

The specific form \eqref{defmatrixa2}
allows, as in the previous subsection, to write the polynomial $HU$ as a sum of positive terms:
\beqa
\label{suma2}
HU_{G,{\bar V}} (t) &=&  \sum_{K}  s^{2g-k_{K}} \ n_{K}^2
\prod_{\ell \in K} \frac{1+t_\ell^2}{2t_\ell} \prod_{\ell' \in \{1,\ldots, L\}} t_{\ell'}\ .
\eeqa

As for the Grosse-Wulkenhaar model, it was proven in \cite{param2} that one
can compute some leading terms. Indeed, 
when
choosing 
$$ K= \{1, \ldots, L\} - J_0, $$
where $J_0$ is an admissible set (as defined in the previous subsection), one has
$$ n_K= 2^g\prod 2 (\Omega \pm 1)$$
(the product of the factors $(\Omega \pm 1)$ depending on the topology of the
graph). As in the previous section one can now set a lower limit on the
polynomial $HU$
\beqa
\label{limit2}
HU_{G, {\bar V}} (t) \ge  \sum_{J_0\, {\rm admissible}}  s^{2[g+(F-1)]} \ 
\left( 2^g \prod 2 (\Omega \pm 1) \right)^2\nonumber\\
\prod_{\ell \in I} \frac{1+t_\ell^2}{2t_\ell} \prod_{\ell' \in \{1,\ldots,L\}} t_{\ell'},
\eeqa
and this is the main result of \cite{param2}. 

\bigskip

The main results \eqref{limit} and \eqref{limit2} allow one to obtain the
following power counting for both these models
$$ \omega = 4g + \frac 12 (N-4),$$
where $\omega$ is the superficial divergence degree and $N$ is the number of external legs of the respective Feynman graph.

Note that for both type of models, similar results of positivity, boundness
and 
power counting 
have been obtained in
\cite{param1} and resp. \cite{param2}  for the second polynomial $HV$ too.

\section{Examples}
\resetequ

In this section we give some examples of the polynomials $HU$. Note that the
root is always chosen to be the vertex denoted by $(x_1,\ldots, x_4)$.

\begin{figure}[ht]
\centerline{\epsfig{figure=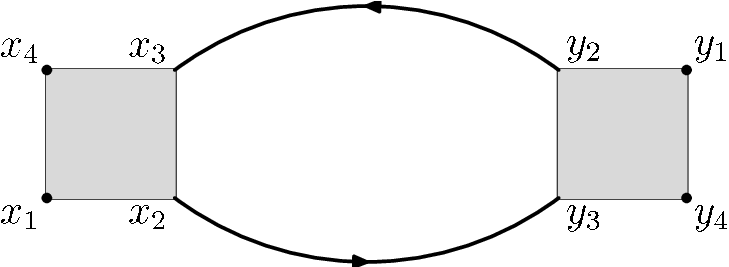,width=6cm}}
\caption{The bubble graph}\label{bubble}
\end{figure}

For the bubble graph (see Fig. \ref{bubble}) one has, in the case of the
Grosse-Wulkenhaar model,
\beqa
\label{pol-bula}
HU_{G,\bar V}&=&(1+4s^2)(t_1+t_2+t_1^2t_2+t_1t_2^2).
\eeqa
In the case of the LSZ model, one has 
\beqa
HU_{G, \bar V}=2 s^2 (t_1+ t_2 + t_1^2t_2+t_1t_2^2)(\Omega -1 )^2. 
\eeqa

\begin{figure}[ht]
\centerline{\epsfig{figure=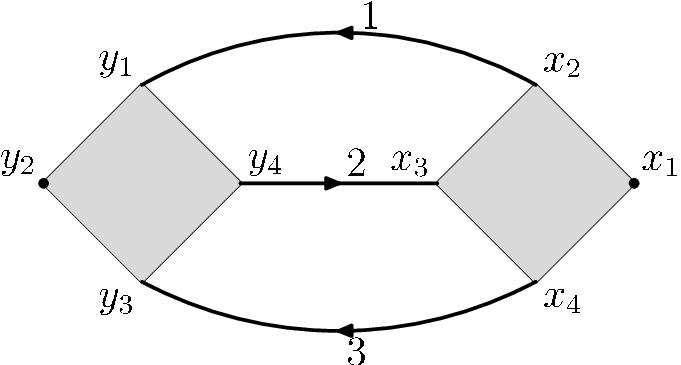,width=6cm}}
\caption{The sunshine graph}\label{sunshine}
\end{figure}

For the sunshine graph (see Fig. \ref{sunshine}) one has, in the case of the
Grosse-Wulkenhaar model,
\beqa
\label{pol-sunshine}
HU_{G,\bar V}&=&\Big{[} t_1t_2+t_1t_3+t_2t_3+t_1^2t_2t_3+t_1t_2^2t_3+t_1t_2t_3^2\Big{]}
(1+8s^2+16s^4)\nonumber\\
&&+16s^2(t_2^2+t_1^2t_3^2).
\eeqa
In the case of the LSZ model, one has:
\beqa
 HU_{G,\bar V}=8s^4(t_1t_2+t_1t_3+t_2t_3+t_1^2t_2t_3+t_1t_2^2t_3+t_1t_2t_3^2)(\Omega -1)^2(\Omega+1)^2.\nonumber\\
\eeqa

We end this section by a more complicated example.

\begin{figure}[ht]
\centerline{\epsfig{figure=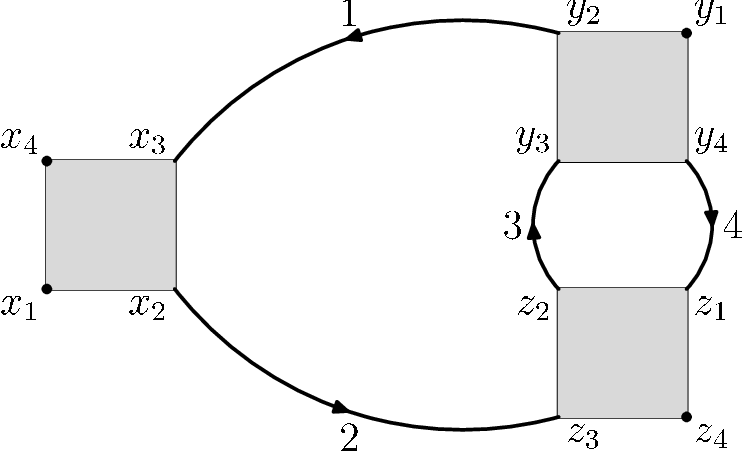,width=6cm}}
\caption{The half-eye Graph}\label{figeye}
\end{figure}

For the Grosse-Wulkenhaar
model, the graph of Fig. \ref{figeye} leads to
\beqa\label{hueye1}
HU_{G,v_1}&=&(A_{24}+A_{14}+A_{23}+A_{13}+A_{12})(1+8s^2+16s^4)\nonumber\\
&&+t_1t_2t_3t_4(8+16s^2+256s^4)+4t_1t_2t_3^2+4t_1t_2t_4^2\nonumber\\
&&+16s^2(t_3^2+t_2^2t_4^2+t_1^2t_4^2+t_1^2t_2^2t_3^2)\nonumber\\
&&+64s^4(t_1t_2t_3^2+t_1t_2t_4^2)\,,
\eeqa
(where for example $A_{24}=t_1t_3+t_1t_3t_2^2+t_1t_3t_4^2+t_1t_3t_2^2t_4^2$). 
For the LSZ model one has 
\beqa
HU_{G, \bar V}=4 s^4 (\Omega-1)^2  (\Omega+1)^2 
[ t_3 t_4 + t_2^2 t_3 t_4 + 
        t_2 (t_3 + t_4 + t_3^2 t_4 + t_3 t_4^2) \nonumber\\
+       t_1^2\left( t_3 t_4 + t_2^2 t_3 t_4 + 
              t_2(t_3 + t_4 + t_3^2 t_4 + t_3 t_4^2)\right) \nonumber\\ 
        + t_1\left( (1 + t_2^2)(t_4 + t_3^2 t_4) + 
              t_3\left( 1 + 64 s^2 t_2 t_4 + t_4^2 + t_2^2(1 + t_4^2)\right)\right)].
\eeqa





\begin{thebibliography}{99}

\bibitem{book-connes}
A. Connes, ``G\'eometrie non commutative'', InterEditions, Paris (1990).


\bibitem{DN}
Douglas M., Nekrasov N.:
Noncommutative field theory.
Rev. Modern Physics 73, 977–1029 (2001). 

\bibitem{Szabo}
R.~J.~Szabo,
  ``Quantum field theory on noncommutative spaces,''
  Phys.\ Rept.\  {\bf 378}, 207 (2003)
  [arXiv:hep-th/0109162].


\bibitem{CDS} 
Connes A, Douglas M. R., Schwarz A.: 
Noncommutative Geometry and Matrix Theory: Compactification on Tori.
JHEP 9802, 3-43 (1998)

\bibitem{SW}
Seiberg N., Witten E.:
String theory and noncommutative geometry.
JHEP { 9909}, 32-131 (1999)  

\bibitem{Suss} 
Susskind L.: The Quantum Hall Fluid and Non-Commutative Chern Simons Theory.

\bibitem{Poly} 
Polychronakos A. P.:
Quantum Hall states on the cylinder as unitary matrix Chern-Simons theory.
JHEP, { 06}, 70-95 (2001)


\bibitem{HellRaams} 
Hellerman S., Van Raamsdonk M.:
Quantum Hall physics equals noncommutative field theory.
JHEP { 10}, 39-51 (2001) 


\bibitem{ls}
E.~Langmann and R.~J.~Szabo,
  ``Duality in scalar field theory on noncommutative phase spaces,''
  Phys.\ Lett.\  B {\bf 533}, 168 (2002)
  [arXiv:hep-th/0202039].


\bibitem{GW1}
Grosse H. and Wulkenhaar R.,
Power-counting theorem for non-local matrix models and renormalization,
 Commun.\ Math.\ Phys. {254}, 91-127 (2005)

\bibitem{GW2} 
Grosse H., Wulkenhaar R., 
Renormalizationof $\phi^4$-theory on noncommutative ${\mathbb R}^4$ in the matrix
base,
 Commun.\ Math.\ Phys. { 256}, 305-374 (2005)

\bibitem{GW3}
Rivasseau V., Vignes-Tourneret F., Wulkenhaar R.:
Renormalization of noncommutative $\phi^{\star4}_4$-theory by multi-scale 
analysis.
Commun. Math. Phys. { 262}, 565-594 (2006)


\bibitem{RVW}
Rivasseau V., Vignes-Tourneret F., Wulkenhaar R.:
Renormalization of noncommutative $\phi^{\star4}_4$-theory by multi-scale 
analysis.
Commun. Math. Phys. { 262}, 565-594 (2006)

\bibitem{GMRV}
Gur{\u{a}}u R., Magnen J., Rivasseau V., Vignes-Tourneret F.:
Renormalization of Non Commutative $\Phi^4_4$ Field Theory in Direct Space.
Commun.\ Math.\ Phys. {267}, 515-542 (2006) 

\bibitem{param1}
R.~Gur\u au and V.~Rivasseau,
  `Parametric representation of noncommutative field theory,''
  Commun.\ Math.\ Phys.\  {\bf 272}, 811 (2007)
  [arXiv:math-ph/0606030].

\bibitem{dimreg}
R.~Gurau and A.~Tanasa,
  ``Dimensional regularization and renormalization of non-commutative QFT,''
  submitted to Annales Henri Poincare, 
  arXiv:0706.1147 [math-ph].

\bibitem{ultimul}
  H.~Grosse and R.~Wulkenhaar,
  ``8D-spectral triple on 4D-Moyal space and the vacuum of noncommutative gauge
  theory,''
  arXiv:0709.0095 [hep-th].


\bibitem{spectral}
  A.~H.~Chamseddine, A.~Connes and M.~Marcolli,
  arXiv:hep-th/0610241.

\bibitem{beta1}
H.~Grosse and R.~Wulkenhaar,
  ``The beta-function in duality-covariant noncommutative phi**4 theory,''
  Eur.\ Phys.\ J.\  C {\bf 35}, 277 (2004)
  [arXiv:hep-th/0402093].

\bibitem{beta23}
M.~Disertori and V.~Rivasseau,
  ``Two and three loops beta function of non commutative phi(4)**4 theory,''
  Eur.\ Phys.\ J.\  C {\bf 50}, 661 (2007)
  [arXiv:hep-th/0610224].

\bibitem{beta}
 M.~Disertori, R.~Gurau, J.~Magnen and V.~Rivasseau,
  ``Vanishing of beta function of non commutative phi(4)**4 theory to all
  orders,''
  Phys.\ Lett.\  B {\bf 649}, 95 (2007)
  [arXiv:hep-th/0612251].

\bibitem{carte} 
Rivasseau V.: From perturbative to Constructive Field Theory: 
Princeton University Press, (1991)

\bibitem{constructiva1}
  V.~Rivasseau,
  ``Constructive Matrix Theory,''
  arXiv:0706.1224 [hep-th].

\bibitem{constructiva2}
  J.~Magnen and V.~Rivasseau,
  ``Constructive $\phi^4$ field theory without tears,''
  arXiv:0706.2457 [math-ph].

\bibitem{goldstone}
A. Marcillaud de Goursac, A. Tanasa,  and
J-C. Wallet, ``Vacuum states of renormalizable non-commutative scalar models;
  on the Goldstone theorem for the linear sigma model'', in progress.

\bibitem{V} 
Vignes-Tourneret F.:
``Renormalization of the orientable non-commutative Gross-Neveu model''.
Ann. Henri Poincar\'e (in press)



\bibitem{lsz}
Langmann E., Szabo R. J., Zarembo K.:
Exact solution of quantum field theory on noncommutative phase spaces.
JHEP { 0401}, 17-87 (2004)

\bibitem{param2}
  V.~Rivasseau and A.~Tanasa,
  ``Parametric representation of 'critical' noncommutative QFT models,''
  arXiv:math-ph/0701034.

\bibitem{sefu}
  V.~Rivasseau,
  ``Non-commutative renormalization,''
  arXiv:0705.0705 [hep-th].

\bibitem{raimar}
  R.~Wulkenhaar,
  ``Field Theories On Deformed Spaces,''
  J.\ Geom.\ Phys.\  {\bf 56}, 108 (2006).

\bibitem{mellin}
  R.~Gurau, A.~P.~C.~Malbouisson, V.~Rivasseau and A.~Tanasa,
  ``Non-Commutative Complete Mellin Representation for Feynman Amplitudes,''
  arXiv:0705.3437 [math-ph].

\bibitem{hopf}
 A.~Tanasa and F.~Vignes-Tourneret,
  ``Hopf algebra of non-commutative field theory,''
submitted to J. Noncomm. Geom.,   
arXiv:0707.4143 [math-ph].

\bibitem{gauge1}
A.~de Goursac, J.~C.~Wallet and R.~Wulkenhaar,
  ``Noncommutative induced gauge theory,''
  arXiv:hep-th/0703075.

\bibitem{gauge2}
H.~Grosse and M.~Wohlgenannt,
  ``Induced Gauge Theory on a Noncommutative Space,''
  arXiv:hep-th/0703169.

\bibitem{gauge3}
D.~N.~Blaschke, H.~Grosse and M.~Schweda,
  ``Non-commutative U(1) Gauge Theory on R**4 with Oscillator Term,''
  arXiv:0705.4205 [hep-th].

\bibitem{gauge4}
  J.~C.~Wallet,
  ``Noncommutative Induced Gauge Theories on Moyal Spaces,''
  arXiv:0708.2471 [hep-th].


\bibitem{gauge5}
A. Marcillaud de Goursac, in progress


\bibitem{itz} 
Itzkinson C., Zuber J.-B.: Quantum Field Theory:
McGraw-Hill, New York (1980)










  
\end{thebibliography}
\end{document}